# AES Analysis of Al Bond Pads in Electrically Insulating Surroundings Utilizing Metal Grid Contacting

*Uwe Scheithauer, ptB, 82008 Unterhaching, Germany*
*E-Mail: scht.ptb@t-online.de; scht.uhg@mail.de*
*Internet: orcid.org/0000-0002-4776-0678; www.researchgate.net/profile/Uwe_Scheithauer*

**Keywords:** AES, Auger electron spectroscopy, bond pad, sputter depth profiling, nonconductive sample

## Abstract:

Bond pads are the electrical interconnections of a microelectronic device to the 'outside world'. A polyimide (PI) layer on top of a microelectronic device protects the whole device against environmental impacts. The bond pads are accessible though openings in this electrically insulating layer. The oxide layer thickness and contaminations at the Al bond pad surface influence as well the quality of the mechanical and electrical joint between bond pad and bond wire as the durability of this interconnection.

If a bond pad surface has to be analyzed with high spatial resolution Auger electron spectroscopy (AES) is the method of choice. AES utilizes an electron beam for excitation, which induces serious sample charging because of the PI layer.

Sample charging can be avoided by metal grids, which are common in transmission electron microscopy (TEM) sample preparation. A TEM grid is pressed onto the sample while the bond pad of interest is centered in the square openings of the grid. Exemplarily analyses of bond pads demonstrate the applicability of this approach for AES measurements.

## 1. Introduction

Auger analysis of insulating materials is often impeded by sample charging. A surface charge severely distorts and shifts secondary electron (SE) spectra, which includes the Auger electron peaks, warping the Auger data meaningless. These effects can also occur when analyzing conductive materials as metal lines or bond pads, which are embedded in insulating materials.

The balanced surface potential is unpredictable for each bond pad analyzed by an AES measurement. This is due to the electrical functionality, to which each bond pad is connected. This functionality is unknown and manifold. In particular the insulating protective top layer of the microelectronic device causes a considerable samples charging. The protective layer of the samples analyzed here is an electrically nonconductive PI, which surrounds the bond pad. The PI above the bond pad is opened by a plasma etching process, which uses fluorinated hydrocarbons.

Bond pads surface contaminations due to the device fabrication process influence the durability of the joint between bond wire and the bond pad. If the thickness of the oxide layer at the Al bond pad surface is too high it impedes a good mechanical and electrical joint between the bond wire and the bond pad. [1] [2]

AES is the method of choice if the bond pad surface has to be analyzed with high spatial resolution. Some kind of charge compensation is necessary to enable reliable AES measurements of bond pads. As demonstrated here a mechanical contacting of the samples by metal grids enables reliable AES measurements on the bond pads of microelectronic samples. This approach requires some experimental skills and experience to handle the sample mounting. But utilizing this approach AES measurements can be done without specific additional instrumental hardware such as low energy ion guns [3] [4].

This paper presents measurements of bond pad surfaces 'as received' and a sputter depth profile measurement. All the measurements are recorded using charge compensation by metal grids.

## 2. Instrumentation

The here presented AES measurements were acquired using Physical Electronics Auger microprobes: a PHI 660 [5] or a PHI 670 [6]. The PHI 660, an instrument with a $LaB_6$ electron emitter, has a resolution of ~ 100 nm under analytical working conditions supplying a sufficiently high primary electron current. The PHI 670 has a hot field emitter electron source and therefore a better lateral resolution of ~ 30 nm under working conditions.

Both instruments are equipped with a differentially pumped sputter ion gun (model 04-303), which is mounted under an angle of 85° with respect to the primary electron beam. The ion gun has a focus of a few 100 µm. The ion beam is rastered electrostatically to obtain laterally homogeneous sputter erosion. The actual sputter rate is monitored by a measurement of the ion beam current using a Faraday cup. For calibration purpose a sputter depth profile measurements of a $SiO_2$ layer of known thickness is used. Since its thickness is known from ellipsometry this depth profile measurement enables to convert the ion beam current measurement into a sputter rate of 'nm $SiO_2$ equivalents per minute' [7].

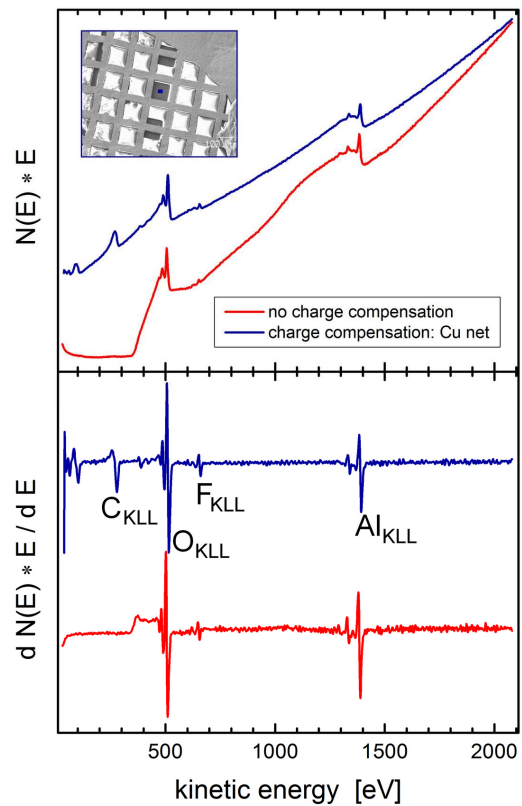

Fig. 1: AES survey scan 'as received'
Al bond pads: ~ 80 x 80 µm$^2$
passivation layer: ~ 6 µm PI

The PHI software program Multipak 6.1 was used for the AES data evaluation [8]. In the case of quantification of measured peak intensities, it uses a simplified model where all detected elements are distributed homogeneously within the analyzed volume. This volume is defined by the analysis area and the information depth of AES, which is derived from the mean free path of electrons [9]. With use of this quantification approach, one monolayer on top of a sample quantifies to ~ 10–30 at% depending on the sample details.

## 3. Experimental Details

Microelectronic device have a protective layer on top. The samples analyzed here are covered by a PI layer, which is a few µm thick. Above the bond pads the PI layer is opened by plasma etching to give access to them. The PI layer is an extremely non-conductive material. Therefore AES measurements at bond pads are almost impossible without adequate charge compensation.

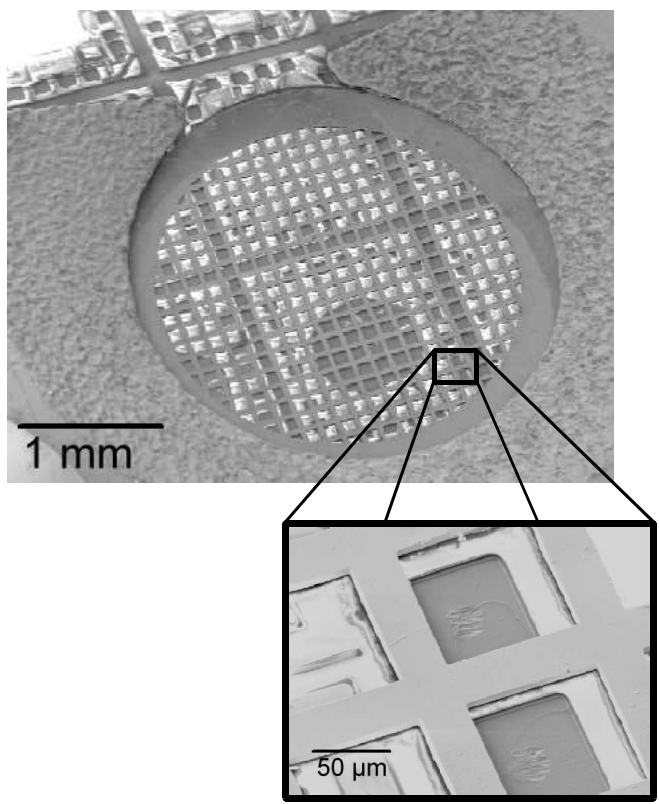

Fig. 2: SE image of the mounted sample with TEM grid Al bond pads: ~ 80 x 80 µm$^2$

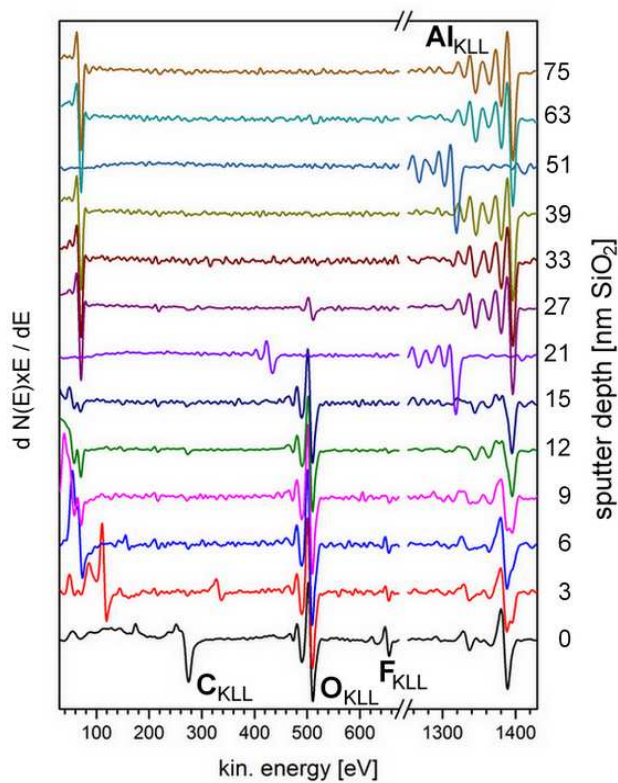

Fig. 3: wide energy range spectra measured during the sputter depth profile

Here square meshed metal grids, which are common for sample mounting in TEM, are used for this purpose. These TEM grids are commercially available with different spacing and different bar widths. They are made of pure metals as Cu, Ni or Au, for instance. Here Cu TEM grids are used.

The TEM grids are placed on the microelectronic devices utilizing an optical microscope. The bond pad of interest is centered in a square opening (insert fig. 1, fig. 2, insert fig 4, fig. 5 and insert fig. 6). A special Ti device holds the TEM grid in position (fig. 2). It has an opening at one side to give the ion gun a good access to the bond pads.

The TEM grid should be pressed as hard as possible onto the PI, which surrounds the bond pad, to achieve good charge compensation. But TEM grids are very thin and flexible, making high pressure mounting difficult. The following approach was proven to be useful: By two tweezers, which have flat and ~ 5 mm wide tips, the TEM grid is bended by an angle of ~ 5° along one symmetry axis. This deformation stiffens the TEM grid along the kink. If this kink area is placed onto the bond pad of interest this guarantees a good mechanical contact to the PI layer.

Fig 1 compares results of AES measurements on bond pads of the same sample without charge compensation and utilizing TEM grid contacting. For the measurements a PHI 670 Auger microprobe was used. The upper part of fig. 1 shows the direct spectra. Without charge compensation undistorted signals could by recorded only for kinetic energies above ~ 350 eV. With TEM grid contacting up to a kinetic energy of ~ 110 eV distortions of the signal are present and therefore the $C_{KLL}$ peak is measured now. The apparent sample surface composition is estimated to be C:O:F:Al = 26:38:2:34 atomic percent.

## 4. Sputter Depth Profile Analysis of a Bond Pad

Fig. 2 shows an SE image of the sample, which is mounted underneath a TEM grid for charge compensation purpose. A SE image of the bond pad of

U. Scheithauer: AES Analysis of Al Bond Pads in Electrically Insulating Surroundings Utilising Metal Grid Contacting

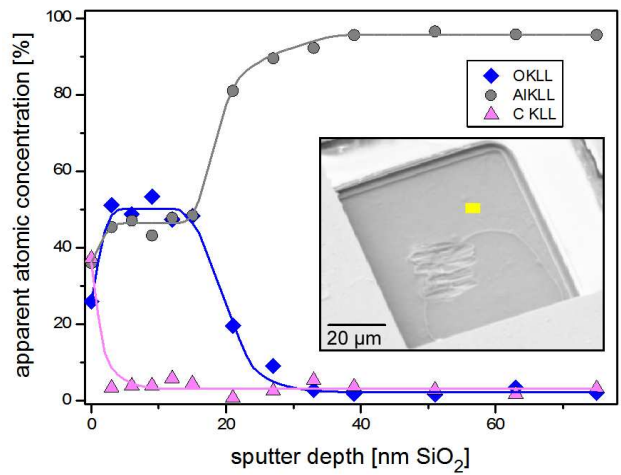

Fig. 4:  AES depth profile measurement
The measurement areas is marked within the inserted SE image.

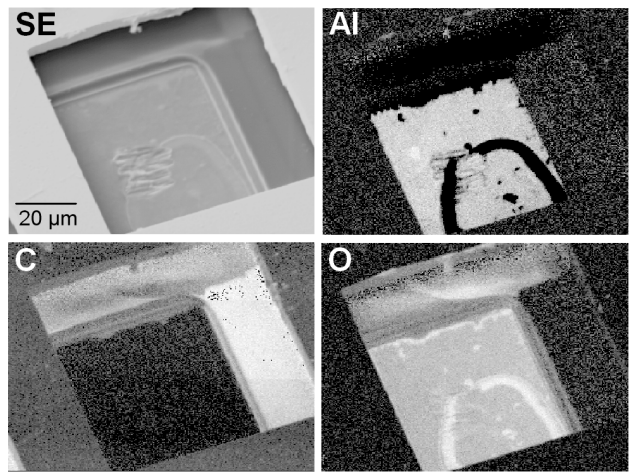

Fig. 5:  SE image and element mappings of C, Al and O

interest at higher magnification is shown at the insert of fig. 4. Mechanical scratches in the bond pads center indicate that the bond pad was contacted by needles for electrical measurements. In the lower part of the bond pad a circular structure is visible.

First, by an AES depth profile measurements the Al surface oxide thickness has to be estimated in regular and inconspicuous area of the bond pad. The analysis area for this depth profile measurement is marked in the insert of fig. 4. Second, the characteristics of the circular structure at the lower part of the bond should be analyzed. Both analytical tasks can only be solved utilizing the spatial resolution of AES.

The measurements were done using the PHI 660 AES microprobe. The sample was tilted by 30° towards the ion guns. Therefore the ion impact angle is ~ 60° relative to the sample normal.

In each sputter depth a wide energy range spectrum was measured during the depth profile measurements [10]. This way the spectra are evaluable even if in a certain sputter a larger peak shift occurs due to surface charging. Fig. 3 shows the spectra measured during the AES sputter depth profile. At the surface 'as received' C, O, F and Al are detected. Al surfaces are always oxidized. F is a residue due to the etching process, which removes the PI above the Al bond pads. And C can be explained by contamination due to sample handling in ambient air.

Distortions are detectable in the low energy region of the spectra up to a kinetic energy of ~ 200 eV and up to a sputter depth of 15 nm $SiO_2$ equivalents. Some signals are identified as spectral distortions, because they have peaks shapes, which are atypical for AES peaks. Whereas most of the spectra are shifted to lower energy by ~ 8 eV, the spectra at a sputter depth of at 21 and 51 nm $SiO_2$ equivalents show a shift of ~ 93 eV towards lower energy. The O and Al signal are identified by the peak shape and the energy distance between the O and Al signal, respectively. For a good signal-to-noise ratio each spectrum shown in fig. 3 is an overlay of several scans over the total kinetic energy range. These single scans were recorded one after the other in a certain sputter depth. The surface potential remains unchanged for the whole measurement time of 10 minutes or longer shown by the constructive overlay of the single scans in fig. 3.

Fig. 4 summarizes the results of the depth profile measurement. At the bond pad surface an Al oxide layer is detected by the peak shape of $Al_{KLL}$ signal and the detection of O. The Al oxide layer is ~ 20 nm $SiO_2$ equivalents thick. Most likely a wire bonding to a bond pad with such a thick oxide layer is nearly impossible.

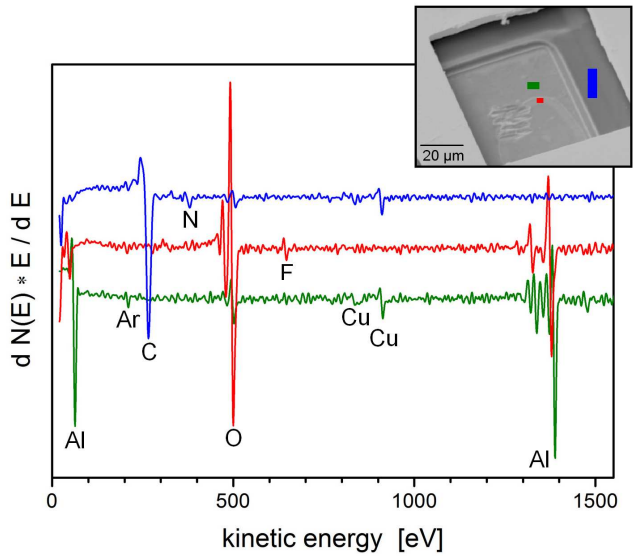

Fig. 6: AES survey scans, measured after sputter depth profile

Fig. 5 shows a SE image and element mappings of the $Al_{KLL}$ signal at metallic peak position, the $C_{KLL}$ signal and the $O_{KLL}$ signal, which are recorded after the depth profile measurement. The sputter ion impact direction comes from the upper left. All images indicate that the sample was not sputtered in the shadow of the TEM grid. On the bond pad a few µm wide circular structure can be indentified. Most likely this is the edge of fluid droplet, which dried on the bond pad. Beside Al oxide an F contamination was detected by an AES measurement of this structure (see fig. 6, red spectrum). Considering this F contamination a reaction mechanism is probable where Al is catalytically converted into its oxide if some moisture is present [11][12] . The main steps of this reaction may be expressed as follows:

$$3\ HF + Al\ \rightarrow AlF_3 + 3/2\ H_2$$
$$AlF_3 + H_2O \rightarrow Al(OH)F_2 + HF$$

This reaction explains the higher oxide layer thickness of the circular structure.

The Cu TEM grid is the highest elevated structure around the analyzed bond pad. This geometrical situation promotes the redeposition of Cu atoms form the TEM grid onto the subjacent microelectronic device areas during the sputter depth profiling. The redeposition occurs especially in forward direction by sputtering of the Cu net edges.

The amount of Cu, which is detected on the different areas of the microelectronic device, depends on the individual composition of these areas and the interaction of the substrate with the impinging $Ar^+$ sputter ions during the sputter process. Cu was detected on the PI and Al surface after the depth profile measurement (fig. 6). The amount of Cu on the PI quantifies to ~ 3 at%. Thanks to this sub monolayer Cu coverage the PI surface becomes conductive and a AES spectrum can be measured [13].

## 5. Conclusion

The charging of a bond pad, which is surrounded by an insulating material, can be avoided during an AES measurement if a metal grid with square openings is pressed onto the microelectronic device and the bond pad is centered in the grid opening. It was demonstrated that utilizing this approach AES spectra of Al bond pad surfaces 'as received' and an AES sputter depth profiles can be recorded successfully.

This sample mounting technique requires some experimental skills and experience. But utilizing this approaches AES measurements of bond pads in insulating surroundings can be done with every AES instrument. No specific additional instrumental hardware is necessary.

**Acknowledgement**

The measurements were done utilizing AES PHI microprobes installed at Siemens AG, Munich, and Infineon, Dresden, Germany. I acknowledge the permission of the Siemens AG to use the measurement results here. For fruitful discussions and suggestions I would like to express my thanks to my colleagues. Special thanks also to Gabi for text editing.

## References

[1] Y.N. Hua, S. Redkar, C.K. Lau, Z.Q. Mo, A Study on Non-Stick Aluminium Bondpads due to Fluorine Contamination using SEM, TEM, IC, Auger, XPS and TOF-SIMS Techniques, Int. Symp. Testing and Failure Anal. (2002) 495-504, https://doi.org/

[2] R.S. Sethu, Reducing non-stick on pad for wire bond: A review, Aust. J. Mechanical Eng. 9 (2012) 147-160, https://doi.org/10.7158/M11-771.2012.9.2

[3] P.E. Larson, M.A. Kelly, Surface charge neutralization of insulating samples in x-ray photoemission spectroscopy, J. Vac. Sci. Technol. A 16 (1998) 3483-3489, https://doi.org/10.1116/1.581507

[4] Physical Electronics Inc., PHI application note, Using Low Energy Ions for Charge Neutralization in PHI Scanning Auger Nanoprobes. Chanhassen, MN 55317, USA (2012)

[5] Physical Electronics Inc., Performance, Engineering and Environmental Specifications: PHI Model 660 Scanning Auger Microprobe. Eden Prairie, MN 55344 USA (1988)

[6] Physical Electronics Inc., Performance, Engineering and Environmental Specificatioms PHI Model 670 Scanning Auger Nanoprobe System. Eden Prairie, MN 55344 USA (1991)

[7] U. Scheithauer, Status monitoring of ion sputter relevant parameters of an XPS depth profiling instrument, Surf. Interface Anal. 52 (2020) 943-947, https://doi.org/10.1002/sia.6765

[8] Physical Electronics Inc., MultiPak Software Manual - Version 6, Eden Prairie, MN 55344 USA (2000)

[9] M.P. Seah, W.A. Dench, Quantitative electron spectroscopy of surfaces: A standard data base for electron inelastic mean free paths in solids, Surf. Interface Anal. 1 (1979) 2-11, https://doi.org/DOI: 10.1002/sia.740010103

[10] U. Scheithauer, The Benefit of Wide Energy Range Spectrum Acquisition During Sputter Depth Profile Measurements, https://arxiv.org/abs/1711.09458 (2017)

[11] U. Scheithauer, Application of the Analytical Methods REM/EDX, AES and SNMS to a Chlorine Induced Aluminium Corrosion, Fresenius J. Anal. Chem. 341 (1991) 445-448, https://doi.org/10.1007/BF00321954

[12] F. Beck (1993) Integrierte Halbleiterschaltungen: Konstruktionsmerkmale, Fehlererscheinungen, Ausfallmechanismen. VCH, Weinheim

[13] U. Scheithauer, Sputter-Induced Cross-Contaminations in Analytical AES and XPS Instrumentation: Utilization of the effect for the In-situ Deposition of Ultra-Thin Functional Layers, Anal. Bioanal. Chem. 405 (2013) 7145–7151, https://doi.org/10.1007/s00216-013-6840-2